\documentclass[epj]{svjour}
\usepackage{graphicx}
\usepackage{amsmath, amssymb}
\begin{document}
\newcommand{\ohbr}{{\hat r}}
\newcommand{\hbr}{{\hat {\bf r}}}
\newcommand{\bryz}{{\bf r}}
\newcommand{\bS}{{\bf s}}
\newcommand{\bSS}{{\bf S}}
\newcommand{\bL}{{\bf L}}
\newcommand{\bD}{{\bf D}}
\newcommand{\bP}{{\bf p}}
\newcommand{\bd}{{\bf d}}
\authorrunning{D. K. Gridnev, S. Schramm, K. A. Gridnev and W. Greiner}
\titlerunning{Instability of neutron matter with 3-body forces}
\title{Nuclear interactions with modern three-body forces lead to the
instability of neutron matter and neutron stars}
\author{Dmitry K. Gridnev\inst{1} \and Stefan Schramm\inst{1} \and Konstantin A. Gridnev\inst{1,2} \and Walter Greiner\inst{1}% etc
% \thanks is optional - remove next line if not needed
%
}                     % Do not remove
\institute{FIAS, Ruth-Moufang-Stra{\ss}e 1, Frankfurt am Main, Germany \and Saint Petersburg State University, Uljanovskaja 1, Saint Petersburg, Russia}
\date{Received: date / Revised version: date}
% The correct dates will be entered by Springer

\abstract{
It is shown that the neutron matter interacting through Argonne V18
pair-potential plus modern variants of Urbana or Illinois three-body forces is
unstable. 
For the energy of $N$ neutrons $E(N)$, which interact through these forces, we
prove mathematically 
that $E(N) = -cN^3 + \mathcal{O}(N^{8/3})$, where $c>0$ is a constant. This
means that: (i) the energy per particle and neutron density diverge rapidly for
large neutron numbers; (ii) bound states of $N$ neutrons exist for $N$ large
enough. The neutron matter collapse is possible due to the form of the repulsive
core in three-body forces, 
which vanishes when three nucleons occupy the same site in space. The old
variant of the forces Urbana VI, where 
the phenomenological repulsive core does not vanish at the origin, resolves this
problem. We prove that to prevent the collapse one should add a repulsive term to the Urbana IX potential, which should be larger than 50 MeV when 3 nucleons 
occupy the same spatial position. 
 % end of PACS codes
} %end of abstract
\maketitle
\section{Introduction}

It is a common place that two liters of water contain twice as much energy as one liter. 
Thermodynamically speaking, this is the result of the energy being an extensive quantity \cite{thirr}. From the quantum mechanical 
point of view this extensivity of the energy can be stated as the following result: let $E(N)$ be the energy of $N$ atoms or molecules, where 
all nuclei and electrons are treated as point particles interacting solely through the Coulomb forces. Then the limit $\lim_{N\to \infty} E(N)/N$ is supposed 
to exist, that is the energy per atom (molecule) approaches a limit in the many-body problem. A formidable task is to prove that the energy can be linearly 
bounded from up and from below $cN \leq E \leq CN$. This type of inequality proves the stability of matter (the most difficult part here 
is to prove the lower bound). 
In their seminal paper \cite{dysonlennard} Dyson and Lennard proved the stability of non-relativistic matter made of pointwise nuclei and electrons, 
see also \cite{liebbook2} on the history of this subject. 
Lieb and Thirring \cite{liebthirring} simplified the argument and improved the value of the constant $c$ by orders of magnitude. 
The corresponding mathematical issues are enlightened in detail in \cite{liebbook2}. The proof of Dyson and Lennard also demonstrated  the 
vital role of the Pauli principle for the stability matter: if electrons were bosons then the energy would not grow linearly in $N$, but rather as 
$E(N) \sim N^{5/3}$, see \cite{liebbook2} for the proof.

Nuclear substance formed by protons and neutrons also forms stable matter. For finite nuclei this is best manifested in the Bethe-Weizs\"acker formula 
\cite{basdevant}. For symmetric nuclear matter (number of protons is equal to the number of neutrons) 
the energy per particle is approximately equal $16$ MeV and the nuclear density is $\rho \simeq 0.16$ fm$^{-3} $\cite{basdevant}. 
Nuclei that are composed solely from neutrons are believed to have positive energy, however, the question of existence of bound state of $N$ 
neutrons, where $N$ is large, is still not 
ultimately resolved \cite{lazauskas,pieper}. Adding to the strong interaction gravitational forces 
enables the creation of neutron stars, which contain about $10^{57}$ neutrons. The astrophysical data regarding masses and radii of these stars 
makes us conclude that the neutrons inside them form neutron matter. 
Let us remark that for stability of nuclear matter the Pauli principle is absolutely essential. 

The basic model of a nuclear system assumes that the Hamiltonian $H = T + \sum_{i< j} v_{ij} + \sum_{i<j<k}v_{ijk}$ 
provides a good description for \textit{any} number of nucleons \cite{panf,urbana}. Here $T$ is the kinetic energy operator, 
$v_{ij}$ and $v_{ijk}$ are two and three-body interactions respectively. There are currently various pair interactions that reproduce 
the available nucleon-nucleon scattering data very well. 
Among those Argonne V18 interaction \cite{av18} is almost local and is well-suited for precise calculations of light nuclei. 
In order to produce correct binding energies of nuclei one has to supplement this two-body potential with the appropriate three-body force. Currently one uses Urbana or 
Illinois three-body interactions \cite{urbana,illinois}. The calculations of light nuclei showed \cite{a1,a2,a3,a4} 
that these interactions represent a highly successful model of nuclear systems,  which predicts very well energy levels and wave functions seen in the experiment. 
Within current models the main attractive part of three-body potentials is contained in the two-pion exchange potential \cite{fuzi,chiral,urbana,illinois}. 
The nuclear matter, however, does not saturate 
satisfactorily with these potentials and one needs a repulsive three-body force \cite{paniold}. This three-body repulsion has a 
phenomenological origin and as we shall see its current form used in Urbana IX and Illinois 7 interactions leads to the collapse of neutron matter: the energy of $N$ 
neutrons behaves like $E(N) \sim -cN^3 + \mathcal{O}(N^{8/3})$, where the constant $c$ is positive. This means that the energy per particle diverges with large $N$. 
Bound multineutrons exist within this model as well, one only needs a large number of neutrons to make them bound. 

Three-body force forms a basic and necessary ingredient of a nuclear force.
The most
direct way to study it is by analyzing scattering data and doing calculations of
light nuclei.
Below we show how a theoretical analysis of neutron matter helps improving the
expression of the 3-body force. The source of the problem leading to the collapse of neutron matter is the fact that 
the 3-body force vanishes when 3 nucleons occupy the same position in space. 
The required corrections are substantial: if one
adds to
Urbana IX a positive Saxon-Woods 3-body term, which prevents instability,
its value at zero, where all 3 nucleons occupy the same spatial position, should be at least 50 MeV. This value is
significant on the scale of 3-body forces. Adding a repulsive term like this
affects
the high density behavior of EOS of neutron matter used in neutron star
calculations. It also changes the overall 3-body force because
with such
repulsive core the attractive part has to be readjusted. Let us stress
that 50
MeV is a minimal required value!
It can get larger depending on the diffuseness of the Saxon-Woods potential and, in
fact,
becomes larger if one considers non-polarized neutron matter. The
repulsive term in the Urbana interaction is isotopically invariant, which
means
that
it appears in normal nuclei, in symmetric nuclear matter etc! Thus the obtained
results affect not only neutron matter, 
they affect all calculations in nuclear systems, which use 3-body forces like Urbana or Illinois.
In Sec.~\ref{sec:3} a simple explanation is given, why the collapse takes place. 

\section{Upper Bound on the Energy of $2N$ Neutrons}

Below we shall construct the upper bound on the energy of interacting neutrons. In the framework of non-relativistic quantum mechanics related bounds 
were obtained in \cite{perez,zhislinvugalter,zhislinnew1,zhislinnew2,seiringer}, where the authors investigated the question of existence of bound states of $N$ identical particles, 
which lie below dissociation thresholds. 
In \cite{zhislinnew1} 
Zhislin has proved the following result. Let  $E(N)$ denote the ground state energy of $N$ fermions (or bosons) that interact through 
the scalar pair potential $v(\bryz)$ satisfying the following condition 
\begin{equation}\label{26:1}
 \int_{\bryz_1, \bryz_2 \in K} v(\bryz_1 - \bryz_2) d \bryz_1 d\bryz_2 < 0
\end{equation}
where $K$ is a fixed arbitrary finite cube in the three-dimensional space. Then $E(N) < -cN^2 $ for $N > N_0$, where $c, N_0 >0$ are constants. 
For $N$ fermions (or bosons) this implies that: (i) for $N_0$ large enough there always exists a negative energy bound state of $N_0$ particles 
irrespectively of a given particle mass; (ii) the particles do not form stable matter, that is the energy per particle diverges if $N \to \infty$. 
The condition Eq.~(\ref{26:1}) can be improved if instead of 
one cubic box in \cite{zhislinnew1,zhislinnew2} one takes two disjoint cubes of equal size $K_1 \cap K_2 = \emptyset$ and requires that 
\begin{gather}
 \int_{\bryz_1, \bryz_2 \in K_1} v(\bryz_1 - \bryz_2) d \bryz_1 d\bryz_2 \nonumber\\
+ \int_{\bryz_1 \in K_1,  \bryz_2 \in K_2 } v(\bryz_1 - \bryz_2) d \bryz_1 d\bryz_2 < 0. \label{26:2}
\end{gather}
Instead of cubes one could use rectangular boxes. We do not prove that the bound $E(N) < -cN^2 $ follows from Eq.~(\ref{26:2}) explicitly, 
but the proof practically mimics the construction below that we use to prove 
the collapse of neutron matter with modern nucleon interactions and is similar to the proof in \cite{zhislinnew2}. 
Note that the condition Eq.~(\ref{26:2}) is fulfilled by the following potential. Suppose that $v(\bryz)$ is continuous, falling off at infinity faster than $|\bryz|^{-2-\delta}$, 
where $\delta >0$ and besides $v(0) = 0$ and $v(\bryz_0) < 0$, where $\bryz_0$ is a fixed three-dimensional vector. To see that Eq.~(\ref{26:2}) with such 
$v(\bryz)$ can be satisfied one can take two cubes $K_{1,2}$ each with the side length $L$, where $K_1$ and $K_2$ have their centers at the origin and 
at $\bryz_0$ respectively. Taking $L$ small enough one ensures that Eq.~(\ref{26:2}) holds. Let us remark that condition Eq.~(\ref{26:2}) is satisfied 
by simplified neutron-neutron pair interactions like Minnesota \cite{minnesota} or Volkov \cite{volkov}. For Minnesota interaction we found a bound multineutron, which contains 
2364 neutrons and has a nuclear density $\rho\simeq 5 \rho_0$ \cite{tobe}.

\begin{figure}
\begin{center}
\includegraphics[width=0.4\textwidth]{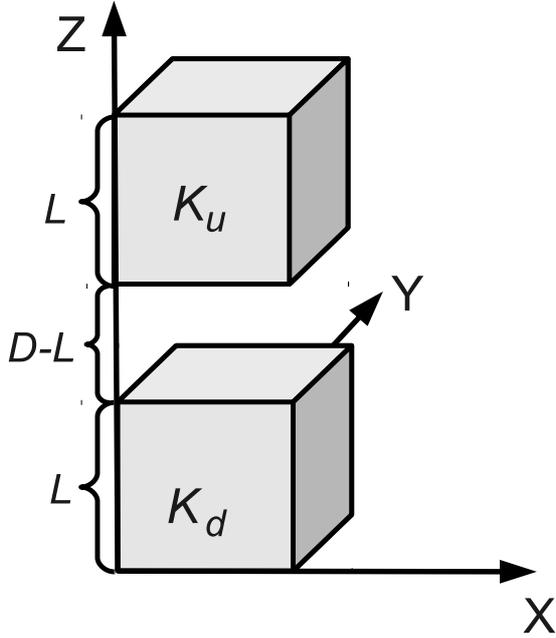}
\caption{The neutrons are placed into two disjoint cubes $K_u, K_d$ each with the side length $L$ (subscripts $u,d$ stand for ``up'' and ``down'' respectively). 
The upper cube is shifted by a distance $D$ along the $Z$-axis with respect to the lower cube.}
\label{fig:1}
\end{center}
\end{figure}

Now let us consider $2N$ neutrons that are described by the following Hamiltonian
\begin{equation} \label{11:4}
H = -\frac{\hbar^2}{2m} \sum_{i=1}^{2N} \Delta_{\bryz_i}  + V_{2b} + V_{3b} =  T + V_{2b} + V_{3b} . 
\end{equation}
The kinetic energy operator $T$ includes the center of mass motion; $m$ is the neutron mass and $\bryz_i$ for $i = 1, \ldots, 2N$ are neutrons' position vectors. 
The term $V_{2b} = \sum_{i<j} v_{ij}$ is the sum of two-nucleon interactions given by the Argonne V18 potential \cite{av18}.  
The term $V_{3b}$ is the three--body interaction, which can be one of the modern versions of Illinois \cite{illinois} or Urbana \cite{urbana} three-nucleon interaction; 
today these are Urbana IX \cite{Urbana IX} and Illinois 7 \cite{illinois7}. The analysis below is restricted to the case of Urbana IX 
interaction, however, the method of analysis and conclusions fully apply to the Illinois 7 interaction as well. 
It is believed \cite{urbana} that Hamiltonian Eq.~(\ref{11:4}) with Urbana or Illinois three-body forces describes 
a stable neutron matter, for the equation of state of neutron matter based on calculations using Eq.~(\ref{11:4}), see \cite{akmal}. 
The main aim of the present paper is to show that neutron matter described by Eq.~(\ref{11:4}) is unstable. 

Below we shall construct the trial function for $2N$ neutrons, which satisfies the Pauli principle and gives the upper bound on 
the ground state energy of $2N$ neutrons described by Eq.~(\ref{11:4}). 
Let us take two cubes $K_u, K_d$ each with the side length $L$ and 
place the first cube at the origin as it is shown in Fig.~\ref{fig:1} and the second over the first one so that the center of $K_u$ is shifted by the distance 
$D$ along the $Z$ axis compared to the center of $K_d$. In order to prevent overlapping of the cubes we require that $D > L$. 
The trial function would depend on three parameters $L, D, \omega >0$, where $\omega$ is an integer. Following \cite{zhislinnew1,zhislinnew2} 
let us first construct $N$ one particle orthogonal 
wave functions. For any $p = 1, 2, \ldots $ and $x \in \mathbb{R}$ we set  
\begin{equation}\label{11:6}
 \varphi_p (x) =  
\left\{
\begin{array}{lr}
(L/2)^{-1/2}  \sin \bigl(2 \pi p L^{-1} \omega x\bigr)  \quad \rm{if} \; \; x \in [0, L],\\
0  \quad \rm{if} \; \; x \notin [0,L]
\end{array}\right.
\end{equation}
Let us fix the an integer $n$ in a way that makes the inequality $n^3 \leq N < (n+1)^3$ hold. For each $t = 1, \ldots, N$ let us choose a triple of 
positive integers $\{t_1, t_2  , t_3\}$ so that $1 \leq t_1, t_2  , t_3 \leq n+1$ and 
\begin{equation}\label{bled}
|t_1 - t'_1| + |t_2 - t'_2| + |t_3 - t'_3| \neq 0  \quad \quad \rm{for} \; \; t \neq t'. 
\end{equation}
That is all $N$ triples should be different (for example, triples \{1,2,8\} and \{1,2,7\} are different).  
Using these triples 
we define the one particle states for $t = 1, \ldots, N$ as follows 
\begin{equation}
 f_t (\bryz) := \varphi_{t_1} (r^x) \varphi_{t_2} (r^y) \varphi_{t_3} (r^z) , 
\end{equation}
where $r^x, r^y, r^z$ are the Cartesian components of the vector $\bryz$. 

Each cube confines $N$ neutrons, which form an excited state of the Fermi gas. Let us set
\begin{gather}
\Psi_\Pi (\bryz_1, \ldots, \bryz_{2N}) :=  [f_1 (\bryz_1) f_2 (\bryz_2) \cdots f_N (\bryz_N)]\nonumber\\
\times [ f_1 (\bryz_{N+1} - {\bf D}) f_2 (\bryz_{N+2} - {\bf D}) \cdots f_N (\bryz_{2N} - {\bf D})], 
\end{gather}
where ${\bf D} := (0,0,D)$ is a three-dimensional vector. 

Let $\mathbb{S}_{2N}$ denote the permutation group for $2N$ particles, whose elements $g \in \mathbb{S}_{2N}$ permute only the spatial coordinates. 
We define the antisymmetrizer $\mathcal{A}$ as 
\begin{equation}\label{11:12}
 \mathcal{A} = \frac 1{(2N)!} \sum_{g \in \mathbb{S}_{2N}} (-1)^{\pi(g)} g , 
\end{equation}
where $\pi(g)$ denotes the parity of the permutation $g$. Now we 
construct the trial function for $2N$ neutrons as 
\begin{equation}\label{26:4}
 \tilde \Psi_A =  \Psi_A (\bryz_1, \ldots, \bryz_{2N}) |n\uparrow\rangle |n\uparrow\rangle \cdots |n\uparrow\rangle , 
\end{equation}
where the spatial part of the wave function is $\Psi_A := \sqrt{(2N)!}\,  \mathcal{A} \Psi_\Pi $. 
 
In Eq.~(\ref{26:4}) $|n\uparrow \rangle$ denotes the isospin-spin state of each nucleon (isospin down for neutron and spin up). 
In the fully polarized trial function we set all neutrons into the state of a neutron with spin up because this simplifies the antisymmetrization. 
It is 
easy to see that $\tilde \Psi_A$ is fully antisymmetric and normalized because the cubes $K_u, K_d$ are disjoint. By the variational principle 
\begin{gather}
 E(2N) \leq \langle \tilde \Psi_A | H | \tilde \Psi_A \rangle 
= \langle \tilde \Psi_A | T | \tilde \Psi_A \rangle + \langle \tilde \Psi_A | V_{2b} | \tilde \Psi_A \rangle \nonumber \\+ \langle \tilde \Psi_A | V_{3b} | \tilde \Psi_A \rangle \label{11:15}
\end{gather}
where $E(2N)$ is the ground state energy of $2N$ neutrons described by Eq.~(\ref{11:4}). 
Let us first consider the contribution from the three-body term. Substituting the interactions from Eqs. (2.1), (2.7) in \cite{urbana} and 
using that both cubes are filled with the same states we get 
\begin{gather}
\langle \tilde \Psi_A | V_{3b} | \tilde \Psi_A  \rangle = \sum_{1 \leq i<j<k \leq 2N} \langle \Psi_A | W(\bryz_i, \bryz_j, \bryz_k)|\Psi_A \rangle \nonumber\\
  =  
\sum_{1 \leq i<j<k \leq 2N} (2N)! \langle \mathcal{A} \Psi_\Pi | W(\bryz_i, \bryz_j, \bryz_k)|\Psi_\Pi \rangle 
= \nonumber \\
2(2N)! \sum_{1 \leq i<j<k \leq N} \langle \mathcal{A}\Psi_\Pi | W(\bryz_i, \bryz_j, \bryz_k)|\Psi_\Pi \rangle \nonumber\\
 + 2(2N)! \sum_{\substack{1 \leq i\leq N \\ N+1\leq j < k \leq 2N}} \langle \mathcal{A}\Psi_\Pi | W(\bryz_i, \bryz_j, \bryz_k)|\Psi_\Pi \rangle . \label{11:16}
\end{gather}
Here 
\begin{gather}
W (\bryz_1, \bryz_2 , \bryz_3) = 
2 A_{2\pi} \sum_{cycl} \Bigl\{ 18 (\hbr_{12} \cdot \hbr_{13} ) \ohbr_{12}^z \ohbr_{13}^z T(r_{12}) T(r_{13}) \nonumber \\
+ 6 (\ohbr_{12}^z )^2 T(r_{12}) \bigl[Y(r_{13}) - T(r_{13})\bigr] 
+ 6 (\ohbr_{13}^z )^2 T(r_{13}) \nonumber\\
\times \bigl[Y(r_{12}) - T(r_{12})\bigr]
+ 2 \bigl[Y(r_{12}) - T(r_{12})\bigr]\nonumber\\
\times \bigl[Y(r_{13}) - T(r_{13})\bigr]\Bigr\} 
+ U_0 \sum_{cycl} T^2 (r_{12}) T^2 (r_{13}) , \label{11:17}
\end{gather}
where $\sum_{cycl}$ denotes a sum over cyclic permutations of the indices $\{1,2,3\}$ and $\bryz_{ik} := \bryz_i - \bryz_k$, $r_{ik} := |\bryz_{ik}|$ and 
$\hbr_{ik} := \bryz_{ik} / |\bryz_{ik}|$. 
The functions $T(r), Y(r)$ are given in Eqs. (2.2), (2.3), (A.1) (A.2) in \cite{urbana}, namely, 
\begin{gather}
Y(r) = \frac{e^{-\mu r}}{\mu r} \bigl[1-e^{-br^2}\bigr], \\
T(r) = \Bigl(1+\frac{3}{\mu r} +\frac{3}{\mu^2 r^2} \Bigr) \frac{e^{-\mu r}}{\mu r} \bigl[1-e^{-br^2}\bigr]^2 , 
\end{gather}
where $\mu = (m_{\pi_0} + 2m_{\pi_{\pm}}) c/(3\hbar)$ is the average of the pion masses and $b = 2.0$ fm$^{-2}$. 
Expanding the exponents it is easy to see that $T(0) = Y(0) = 0$, which means that the whole three-body interaction vanishes if three nucleons occupy the same position 
in space. For Urbana IX interaction the values of the constants appearing in Eq.~(\ref{11:17}) are $A_{2\pi} = -0.0293$ MeV and $U_0 = 0.0048$ MeV. 

\begin{figure}
\begin{center}
\includegraphics[width=0.5\textwidth]{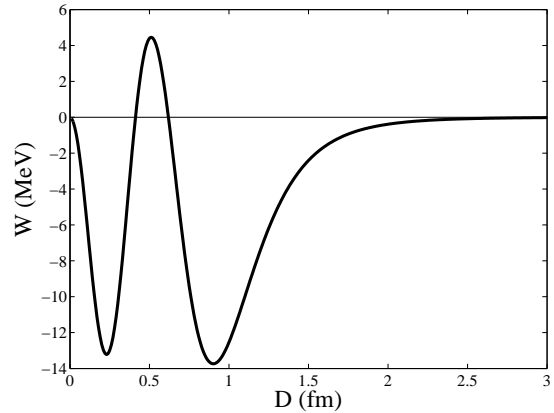}
\caption{The plot of the function $W(0,0,{\bf D})$ (where ${\bf D} \equiv (0,0,D)$) versus parameter $D$.}
\label{fig:2}
\end{center}
\end{figure}

Now we fix the parameters $L, D$ of the trial function in the following way. 
Let us define the constants 
\begin{gather}
  B_1 := \int_{\bryz_1, \bryz_2, \bryz_3 \in K_d} W (\bryz_1, \bryz_2 , \bryz_3) d\bryz_1 d\bryz_2 d\bryz_3  \label{11:20}\\
  B_2 := \int_{\bryz_1 \in K_d \atop \bryz_2, \bryz_3 \in K_u} W (\bryz_1, \bryz_2 , \bryz_3) d\bryz_1 d\bryz_2 d\bryz_3 \nonumber\\= 
\int_{\bryz_1, \bryz_2, \bryz_3 \in K_d} W(\bryz_1, \bryz_2 + {\bf D}, \bryz_3+ {\bf D}) d\bryz_1 d\bryz_2 d\bryz_3 . \label{11:21}
\end{gather}
We set the values of $L, D$ so as to make the following inequality holds 
\begin{equation}\label{11:22}
 Q := -(B_2 + (1/3)B_1) >0 . 
\end{equation}
To see that Eq.~(\ref{11:22}) can be fulfilled it is not necessary to calculate the integrals in Eqs.~(\ref{11:20})-(\ref{11:21}) numerically. Note that $W (\bryz_1, \bryz_2 , \bryz_3)$ is a continuous function. 
For $L$ small enough the integrand in Eq.~(\ref{11:20}) would be close to $W(0,0,0) = 0$, whereas the integrand on the right-hand side of Eq.~(\ref{11:21}) would be close to 
$W(0,0,{\bf D})$. The graph of the function $W(0,0,{\bf D})$, which depends on $D$ is shown in Fig.~\ref{fig:2}. One can set $D = 1$ fm in order to ensure that the integrand in Eq.~(\ref{11:21}) 
would be less than $-12$ MeV. Therefore $Q >0$ if $L$ is fixed sufficiently small though different from zero. 

It remains to fix the last parameter $\omega$ in the trial function. For convenience of notation 
let us introduce the nine-dimensional vector $\bS = (\bryz_1, \bryz_2 , \bryz_3)$ so that $\bS \in \mathbb{R}^9$ has the 
components $\bS = (s_1, \ldots, s_9)=(r_1^x , r_1^y, \ldots, r_3^y , r_3^z)$. Let $\mathcal{D} \subset \mathbb{R}^9$ denote the subset of all \textit{non-zero} 
vectors with integer coordinates (that is for any $\bd \in \mathcal{D} $ all $d_i$ for $i =1, \ldots, 9$ are integers and $\sum_{i=1}^9 d_i^2 \neq 0$). And let us define 
\begin{gather}
\Upsilon_1 (\omega) =  \sup_{\bd \in \mathcal{D}} \Bigl| \rm{Re}\; \int_0^L ds_1 \ldots \int_0^L ds_9  W(\bS)  \nonumber\\
\times \exp \bigl(i 2 \pi L^{-1} \omega (\bd \cdot \bS)\bigr) \Bigr| . \label{11:23}
\end{gather}
Note that $W(\bryz_1, \bryz_2 , \bryz_3)$ is square integrable in the cube, that is 
\begin{equation}
\int_0^L ds_1 \ldots \int_0^L ds_9  \bigl| W(\bS)\bigr|^2 < \infty . 
\end{equation}
It is a trivial consequence of the Bessel's inequality 
that $\Upsilon_1 (\omega) \to 0$ for $\omega \to \infty$, since the integral in Eq.~(\ref{11:23}) is proportional to the Fourier coefficient of the function $W(\bS)$. 
Similarly, we define 
\begin{gather}
\Upsilon_2 (\omega) = \sup_{\bd \in \mathcal{D}} \Bigl| \rm{Re}\; \int_0^L ds_1 \ldots \int_0^L ds_9  \tilde W (\bS) \nonumber\\
\times \exp \bigl(i 2 \pi L^{-1} \omega (\bd \cdot \bS)\bigr) \Bigr|, 
\end{gather}
where by definition $\tilde W(\bS) = W(\bryz_1 , \bryz_2 + \bD , \bryz_3 + \bD)$, and $\Upsilon_2 (\omega) \to 0$ for $\omega \to \infty$ as well. 
Therefore, we can fix $\omega$ requiring that 
\begin{equation}
 (3^9 + 2^9 )\bigl[\Upsilon_1(\omega) + \Upsilon_2 (\omega)\bigr] \leq Q/2, \label{11:26}
\end{equation}
where $Q$ is defined in Eq.~(\ref{11:22}).

Now we turn back to Eq.~(\ref{11:16}). On the right-hand side (rhs) of Eq.~(\ref{11:16}) the antisymmetrization operator Eq.~(\ref{11:12}) enters two times. 
It is easy to check that in the first term on the rhs of 
Eq.~(\ref{11:16}) only 6 permutations $g$ in Eq.~(\ref{11:12}) produce a non-vanishing contribution. These are 6 permutations, which permute the indices $\{i,j,k\}$ and leave all other 
$2N-3$ indices unaffected. By the same arguments in the second term on the rhs of Eq.~(\ref{11:16}) only 2 permutations make non-vanishing contributions, these permute the indices $j,k$ and 
do not permute the other $2N-2$ indices. Thus Eq.~(\ref{11:16}) can be rewritten as 
\begin{gather}
   \langle \tilde \Psi_A | V_{3b} | \tilde \Psi_A  \rangle = 
2 \sum_{1 \leq i<j<k \leq N} \sum_{g \in \mathbb{S}_{\{i,j,k\}}} (-1)^{\pi(g)} \nonumber\\
\times \int_{\bryz_{i,j,k} \in K_d} f_i (\bryz_i) f_j (\bryz_j) f_k (\bryz_k) \nonumber\\
 \times  W(\bryz_i, \bryz_j, \bryz_k) f_{g(i)} (\bryz_i) f_{g(j)} (\bryz_j) f_{g(k)} (\bryz_k) d\bryz_i d\bryz_j d\bryz_k \nonumber\\
 + 2  \sum_{1 \leq i\leq N \atop N+1\leq j < k \leq 2N} \int_{\bryz_{i,j,k} \in K_d}  f_i^2(\bryz_i) 
\bigl[(f_j (\bryz_j ) f_k(\bryz_k ))^2 \nonumber\\- f_j (\bryz_j ) f_k (\bryz_k ) f_j (\bryz_k ) f_k (\bryz_j ) \bigr] \nonumber\\
\times W(\bryz_i, \bryz_j + \bD, \bryz_k + \bD) d\bryz_i d\bryz_j d\bryz_k \label{11:27}
\end{gather}
 In Eq.~(\ref{11:27}) $\mathbb{S}_{\{i,j,k\}}$ denotes the permutation group of the indices $\{i,j,k\}$, which consists of 6 permutations. 
We focus on the first integral on the rhs of Eq.~(\ref{11:27}). Let us first consider the identical permutation $g =1$, that is we set 
$g(i)=i$, $g(j)=j$, $g(k)=k$. The integral in the considered term for $g = 1$ equals
\begin{gather}
 \int_{\bryz_{i,j,k} \in K_d} f^2_i (\bryz_i) f^2_j (\bryz_j) f^2_k (\bryz_k) W(\bryz_i, \bryz_j, \bryz_k) d\bryz_i d\bryz_j d\bryz_k \nonumber\\
 = (2L)^{-9} \rm{Re}\;  \int_0^L ds_1 \cdots \int_0^L ds_9 W(\bS) \nonumber\\
\times [2 - e^{4\pi i i_1 L^{-1}\omega s_1} - e^{-4\pi i i_1L^{-1}\omega s_1}] \nonumber\\
 \times [2 - e^{4\pi i i_2 L^{-1}\omega s_2} - e^{-4\pi i i_2 L^{-1}\omega s_2}]
\times \nonumber\\
\cdots \times [2 - e^{4\pi i k_3 L^{-1}\omega s_9} - e^{-4\pi i k_3 L^{-1}\omega s_9}] , \label{11:28}
\end{gather}
where we have used the explicit expression for one-particle wave functions Eq.~(\ref{11:6}). Now we expand the product of terms in square brackets and 
use Eq.~(\ref{11:20}) and Eq.~(\ref{11:23}) to obtain the upper bound 
\begin{gather}
 \int_{\bryz_{i,j,k} \in K_d} f^2_i (\bryz_i) f^2_j (\bryz_j) f^2_k (\bryz_k) W(\bryz_i, \bryz_j, \bryz_k) 
d\bryz_i d\bryz_j d\bryz_k \nonumber\\
\leq L^{-9}B_1 + (3^9 -1)L^{-9}\Upsilon_1 (\omega)
\end{gather}
Indeed, after expanding the product in Eq.~(\ref{11:28}) we would obtain $3^9$ terms. $3^9-1$ terms would contain at least one exponent function with a non-zero 
argument and thus each of 
these terms can be estimated using Eq.~(\ref{11:23}). 

In a similar way we estimate other terms in Eq.~(\ref{11:27}), which correspond to permutations $g \neq 1$. 
Substituting the explicit expressions for one-particle wave functions we get 
\begin{gather}
 (-1)^{\pi(g)}\int_{\bryz_{i,j,k} \in K_d} f_i (\bryz_i) f_j (\bryz_j) f_k (\bryz_k) W(\bS) \nonumber\\
\times f_{g(i)} (\bryz_i) f_{g(j)} (\bryz_j) f_{g(k)} (\bryz_k) d\bryz_i d\bryz_j d\bryz_k \nonumber\\
 = (-1)^{\pi(g)} (-2L)^{-9} \rm{Re}\;  \int_0^L ds_1 \cdots \int_0^L ds_9 W(\bS) \nonumber\\ 
\times [e^{2\pi i i_1 L^{-1}\omega s_1} - e^{-2\pi i i_1L^{-1}\omega s_1}]  \times \cdots\nonumber\\
 \times [e^{2\pi i g_2(k) L^{-1}\omega s_8} - e^{-2\pi i g_2(k) L^{-1}\omega s_8}] \nonumber \\
\times [e^{2\pi i  g_3(k) L^{-1}\omega s_9} - e^{-2\pi i g_3(k) L^{-1}\omega s_9}]  . 
\label{12:1}
\end{gather}
Because $g\neq 1$ we have that either $g(i) \neq i$ or $g(j) \neq j$. Without loosing generality we can assume that $g(i) \neq i$ and, hence, 
$|g_1 (i) - i_1|+ |g_2 (i) - i_2|+ |g_3 (i) - i_3| \neq 0$. Again, without loosing generality let us assume that $g_1 (i) \neq i_1$. After expanding the product of square brackets in (\ref{12:1}) we would obtain $2^{18}$ terms, each 
of them would contain one of the four terms: $\exp \bigl\{ 2\pi i L^{-1} \omega[i_1 \pm g_1(i)]s_1\bigr\}$ or $\exp \bigl\{ -2\pi i L^{-1} \omega[i_1 \pm g_1(i)]s_1\bigr\}$. In 
none of the four cases the argument of the exponent vanishes, which due to Eq.~(\ref{11:23}) leads to the upper bound 
\begin{gather}
 (-1)^{\pi(g)}\int_{\bryz_{i,j,k} \in K_d} f_i (\bryz_i) f_j (\bryz_j) f_k (\bryz_k) W(\bS) 
 f_{g(i)} (\bryz_i) \nonumber \\\times f_{g(j)} (\bryz_j) f_{g(k)} (\bryz_k) d\bryz_i d\bryz_j d\bryz_k 
\leq 2^{9} L^{-9} \Upsilon_1 (\omega) . \label{12:3}
\end{gather}
We use the same method to estimate the integrals in the last term in Eq.~(\ref{11:27}), which results in the upper bounds 
\begin{gather}
  \int_{\bryz_{i,j,k} \in K_d}  f_i^2(\bryz_i) f_j^2 (\bryz_j ) f_k^2 (\bryz_k ) \tilde W(\bS) d\bryz_i d\bryz_j d\bryz_k  \nonumber\\
\leq L^{-9} B_2 + (3^9 -1) L^{-9} \Upsilon_2 (\omega)
\end{gather}
and 
\begin{gather}
  -\int_{\bryz_{i,j,k} \in K_d}  f_i^2(\bryz_i) f_j (\bryz_j ) f_k (\bryz_k ) f_j (\bryz_k ) f_k (\bryz_j ) \nonumber\\ 
\times \tilde W(\bS) d\bryz_i d\bryz_j d\bryz_k \leq 2^{9} L^{-9} \Upsilon_2 (\omega) . \label{11:32}
\end{gather}
Finally, using Eqs.~(\ref{11:27})-(\ref{11:32}) we estimate the contribution of the three-body term as follows 
\begin{gather}
\langle \tilde \Psi_A | V_{3b} | \tilde \Psi_A \rangle \leq 
2L^{-9} \nonumber\\
\times \sum_{1 \leq i<j<k \leq N} \left\{ B_1 + (3^9 + 5\times 2^9 -1) \Upsilon_1 (\omega) \right\} \nonumber\\
+ 2L^{-9} \sum_{1 \leq i\leq N \atop N+1\leq j < k \leq 2N} \left\{ B_2 + (3^9 + 2^9 -1) \Upsilon_2 (\omega) \right\}  
= 2L^{-9} \nonumber\\\times  \left[ \left\{ B_1 + (3^9 + 5\times 2^9 -1) \Upsilon_1 (\omega) \right\} \frac{N(N-1)(N-2)}{6} \right. \nonumber \\ 
\left. + \left\{ B_2 + (3^9 + 2^9 -1) \Upsilon_2 (\omega) \right\} \frac{N^2 (N-1)}{2} \right] . 
\end{gather}
Now using Eq.~(\ref{11:26}) we get 
\begin{equation}
\langle \tilde \Psi_A | V_{3b} | \tilde \Psi_A \rangle \leq -QL^{-9} N^3 + \mathcal{O} (N^2) \label{11:34}
\end{equation}
The contribution from the kinetic energy term can be estimated as follows
\begin{gather}
   \langle \tilde \Psi_A | T | \tilde \Psi_A \rangle = \langle \Psi_A | T | \Psi_A \rangle = (2N)! \langle \mathcal{A} \Psi_\Pi | T | \Psi_\Pi \rangle \nonumber\\
= 
\langle  \Psi_\Pi | T | \Psi_\Pi \rangle 
 \leq 2\bigl(2\pi \omega L^{-1}\bigr)^2 \sum_{i=1}^{n+1} \sum_{j=1}^{n+1} \sum_{k=1}^{n+1} (i^2 +j^2 + k^2) \nonumber\\= 
\bigl(2\pi \omega L^{-1}\bigr)^2 (n+1)^3 (n+2) (2n+3) \nonumber \\
 \leq \bigl(2\pi \omega L^{-1}\bigr)^2 (N^{1/3}+1)^3 (N^{1/3}+2) \nonumber\\\times (2N^{1/3}+3)  = \mathcal{O} (N^{5/3}) . \label{11:35}
\end{gather}
We have used that in the sum in Eq.~(\ref{11:12}), which enters the expression $\langle \mathcal{A} \Psi_\Pi | T | \Psi_\Pi \rangle$, only 
$g = 1$  produces a non-vanishing contribution. It remains to consider the contribution from $V_{2b}$. 
Providing a very rough upper bound on $\langle \tilde \Psi_a | V_{2b} | \tilde \Psi_A \rangle $ we 
shall prove that this term contributes in Eq.~(\ref{11:15}) as $\mathcal{O} (N^{8/3})$. 
In this case due to Eqs.~(\ref{11:34}), (\ref{11:35}) we would have 
\begin{equation}
 E(2N) < -QL^{-9} N^{3} + \mathcal{O} (N^{8/3}) , \label{11:36}
\end{equation}
 which proves 
the instability of neutron matter: the energy per particle $E(2N) / (2N)$ diverges for large $N$ at least as fast as $N^2$. The Argonne $v_{18}$ pair potential \cite{av18} contains 18 operators and the functions in front of these operators are finite. 
Thus all two-body interactions except the terms containing the operators $\bL_{ij}^2$, $(\bL_{ij} \cdot \bSS_{ij})^2$ and 
$(\bL_{ij} \cdot \bSS_{ij})$ can be bounded by a constant $V_0'$ and contribute to $\langle \tilde \Psi_a | V_{2b} | \tilde \Psi_A \rangle $ as $V_0' N(2N-1) = \mathcal{O}(N^2)$. 
To derive an upper bound on $\langle \tilde \Psi_a | V_{2b} | \tilde \Psi_A \rangle $ it suffices to consider the contribution from the interaction 
$V_0 \sum_{i < j} \bL_{ij}^2$, where $V_0 >0$ is a constant equal to the maximum of the function in front of the $\bL_{ij}^2$ term. 
Other interactions containing the terms $(\bL_{ij} \cdot \bSS_{ij})^2$ and 
$(\bL_{ij} \cdot \bSS_{ij})$ can be considered similarly. The relative orbital angular momentum of the particles $i,j$ is 
$\bL_{ij} = (1/2)(\bryz_j - \bryz_i) \times (\bP_j - \bP_i)$, where $\bP_i = -i\hbar \nabla_{\bryz_i}$. The square of its z-component can be estimated as follows 
\begin{gather}
 \left( L_{ij}^z\right)^2 = \frac 14 \left[(r_j^x- r_i^x)(p_j^y- p_i^y) - (r_j^y- r_i^y)(p_j^x- p_i^x)\right]^2 \nonumber\\
\leq \frac 12 \left[(r_j^x- r_i^x)^2 (p_j^y- p_i^y)^2 + (r_j^y- r_i^y)^2 (p_j^x- p_i^x)^2 \right] \nonumber\\
\leq (r_j^x- r_i^x)^2 \bigl((p_j^y)^2 + (p_i^y)^2\bigr) + (r_j^y- r_i^y)^2 \bigl((p_j^x)^2 + (p_i^x)^2\bigr) . \label{11:37}
\end{gather}
\textit{Remark}\\
The inequalities in Eq.~(\ref{11:37}) are understood as operator inequalities. For self-adjoint operators $A, B$ the inequality $A \leq B$ means that 
$\langle f | A | f \rangle \leq \langle f | B | f \rangle$ for all admissible $f$. In Eq.~(\ref{11:37}) we have used the operator inequality 
$(A-B)^2 \leq 2A^2 + 2B^2$, which easily follows from the obvious $(A+B)^2 \geq 0$. \\
Similarly estimating the squares of x,y-components of $\bL_{ij}$ we finally obtain 
\begin{gather}
 V_0 \langle \tilde\Psi_A | \sum_{i< j} \bL_{ij}^2 | \tilde\Psi_A \rangle  \leq 2(L+D)^2 V_0 \langle \tilde\Psi_A | \sum_{i< j} \bigl( \bP_i^2 + \bP_j^2 \bigr) | \tilde\Psi_A \rangle \nonumber \\
\leq  2(L+D)^2 V_0 (2m) (2N) \langle  \tilde\Psi_A | T| \tilde\Psi_A \rangle = \mathcal{O} (N^{8/3}), 
\end{gather}
where we have used Eq.~(\ref{11:35}) and the fact that spatial coordinates are bounded by the dimensions of the cubes $K_u , K_d$. 
To be consistent, let us consider the interaction terms that contain spin-orbit-squared and spin-orbit terms. Using that the trial function is 
fully polarized (all spins are up) the contribution from 
spin-orbit-squared terms can be bounded as 
\begin{gather}
  U_0 \sum_{i< j} \langle \tilde\Psi_A | (\bL_{ij} \cdot \bSS_{ij})^2 | \tilde\Psi_A \rangle \nonumber\\
=   U_0 \sum_{i< j} 
\langle \tilde\Psi_A | \left( L_{ij}^z\right)^2 + \left( L_{ij}^x S_{ij}^x + L_{ij}^y S_{ij}^y\right)^2| \tilde\Psi_A \rangle \nonumber \\
\leq  U_0 \sum_{i< j} 
 \langle \tilde\Psi_A |  \left( L_{ij}^z\right)^2 + 2\left( L_{ij}^x S_{ij}^x \right)^2+ 2\left( L_{ij}^y S_{ij}^y \right)^2| \tilde\Psi_A \rangle \nonumber\\
= U_0 \sum_{i< j}  \langle \tilde\Psi_A | \bL_{ij}^2 | \tilde\Psi_A \rangle = \mathcal{O} (N^{8/3}), 
\end{gather}
where $U_0$ is a positive constant equal to the maximum of the function in front of the $(\bL_{ij} \cdot \bSS_{ij})^2$ term. Similarly the contribution from 
spin-orbit terms can be bounded as follows 
\begin{gather}
  U'_0 \sum_{i< j} \left| \langle \tilde\Psi_A | (\bL_{ij} \cdot \bSS_{ij}) | \tilde\Psi_A \rangle \right| \leq   U'_0 \sum_{i< j} 
\bigl( \langle \tilde\Psi_A | \bL_{ij}^2 | \tilde\Psi_A \rangle\bigr)^{\frac 12} \nonumber \\
\leq U'_0 \Bigl( 
\langle \tilde\Psi_A | \sum_{i< j} \bL_{ij}^2 | \tilde\Psi_A \rangle\Bigr)^{\frac 12} (N(2N-1))^{\frac 12} = \mathcal{O} (N^{\frac 73}), \label{hagffds}
\end{gather}
where $U'_0$ is a positive constant equal to the maximal absolute value of the function in front of the $(\bL_{ij} \cdot \bSS_{ij})$ term. In (\ref{hagffds}) we have 
used twice the Cauchy-Schwarz inequality. Summarizing, we find that Eq.~(\ref{11:36}) holds. 
Preliminary estimates \cite{tobe} with AV6' potential \cite{av6} instead of AV18 show that a bound multineutron 
with negative energy contains less than 4500 neutrons.  
We also come to the conclusion that the matter-like state of $N$ neutrons for $N$ large is quasistable with modern forces (under matter-like we mean 
the state described in \cite{akmal}). However, with large $N$ ($N \gtrapprox 1000$) there appears another deep energy minimum, which is structurally very different 
from the matter-like state. This minimum is unphysical because it leads to densities, which grow to infinity with large $N$. 
In the subsequent analysis one should analyze the probability of transition into unphysical minimum. 
But necessary changes in the repulsive core of 3-body  forces can be easily introduced in order for this unphysical minimum to disappear. 
In particular, Urbana VI three-body force does not create such unphysical minimum.

A few remarks are in order. Taking a fully polarized trial function  in
(\ref{26:4}) was merely a technical simplification, which
allowed treating antisymmetry of a trial function in a more lucid fashion. An
non-polarized trial function leads to even larger estimate of $|E(2N)|$! 
Another technical trick is to set neutrons into a highly dense state in the trial
function. From the proof, which uses the variational principle, one may get a false impression that we apply the 
modern nuclear Hamiltonian to a media with extreme
densities, which do not occur in Nature. 
This is, however, not true. Instead we prove that ultra high densities inevitably result when $N$ neutrons
are in
the ground state and $N$ is large! We do not claim that such densities appear in Nature. On
the
contrary,
we claim that the force has to be corrected in order to avoid the
appearance of
unphysical densities. That is we demonstrate mathematically that the
repulsive core in the Urbana and Illinois 3-body interactions is wrong, since it
leads to the
growth of the binding energy according to the law of $N^3$. 

It is important to show that the corrections of the 3-body interaction that are required for stability of neutron matter 
are substantial. For that let us consider a hypothetical change in the Urbana IX interaction, which can be written as 
\begin{equation}\label{17.1;1}
 \tilde V_{3b} = V_{3b} + V_{3NR}, 
\end{equation}
where $V_{3b}$ is defined as above and $V_{3NR}$ has the form of the 
repulsive term that is used in Urbana VI interaction, see Eqs.~(2.8)-(2.10) in \cite{urbana}. That is 
\begin{equation} \label{betax2}
 V_{3NR} (\bryz_1, \bryz_2 , \bryz_3) = U_C\mathcal{W} (r_{12})\mathcal{W} (r_{13})\mathcal{W} (r_{23}) 
\end{equation}
where $U_C$ is a constant and 
\begin{equation}
 \mathcal{W} (r) = \left[1 + \exp\bigl((r-R)c^{-1}\bigr)\right]^{-1} 
\end{equation}
with $R = 0.5$ fm and $c = 0.2$ fm. The necessary stability condition in the integral form, which is derived from (\ref{11:22}), reads 
\begin{gather}
 \frac 13 B_1 + B_2 + \frac 13 \int_{\bryz_1, \bryz_2, \bryz_3 \in K_d} V_{3NR} (\bryz_1, \bryz_2 , \bryz_3) d\bryz_1 d\bryz_2 d\bryz_3 \nonumber\\
+ \int_{\bryz_1, \bryz_2, \bryz_3 \in K_d} V_{3NR}(\bryz_1, \bryz_2 + {\bf D}, \bryz_3+ {\bf D}) d\bryz_1 d\bryz_2 d\bryz_3 \geq 0, \label{betax}
\end{gather}
where integrals $B_{1,2}$ are given in Eqs. (\ref{11:20}) and (\ref{11:21}) respectively. 
Inequality (\ref{betax}) should hold for all values of the constants $L, D >0$ such that $D-L >0$ 
(the cubes in Fig.~\ref{fig:1} should remain disjoint). Let us set $D = 0.9$ fm so that $W(0,0,\bD) \simeq -13.7$ MeV, see Fig.~\ref{fig:2}. Taking $L \to 0$ we immediately obtain from 
(\ref{betax}) the minimal value for the constant $U_C$ in (\ref{betax2}), which is $U_C > 51$ MeV. Varying the sizes of the cubes one may obtain a better value and the minimal value of $U_C$ indeed becomes larger if one constructs an non-polarized trial function.

Stability, radius and masses of neutron stars are largely governed by the
equation of state (EOS) of nuclear matter, see f. e.  \cite{akmal,gandol}. 
The present result shows a dramatic effect of the repulsive core on the EOS at
high densities. Let us note that recently there were calculations of neutron
matter \cite{chiral} 
with potentials derived from chiral perturbation theory
\cite{machleidt,epelbaum}. We were not able to reach the conclusions, whether
such instability occurs for such 
interactions; it is important to generalize the stability condition derived in
this paper to interactions in momentum space like in \cite{machleidt}.

\section{Summary}\label{sec:3}

It is a standard practice to
study the neutron matter by considering the ground state of N neutrons, which are set in an
external trap \cite{maris}.
We prove that the neutrons would collapse with large
$N$ if one uses modern 3-body forces. The mathematical proof is absolutely rigorous, the collapse is derived from the Schr\"odinger equation. 
The reason for the collapse is the presence of form-factors in the interactions, which make 3-body force vanish when 3 nucleons occupy the same position is space. 
The neutron density is most probably zero for $N < 100$ (100 neutrons 
seem to have no bound states). As $N$ increases the first bound state of $N$ neutrons emerges at some point and the density starts growing with $N$. 
This happens without any external compression and it is a mathematical fact. By using the mathematical approach we come to the conclusion that 
in order for the modern nuclear Hamiltonian to work one should constrain the number of particles. Otherwise for $N \gtrapprox 1000$ unphysical effects 
begin to dominate. This problem is easily cured by changing the phenomenological repulsive core of the 3-body force, for example, Urbana VI interaction 
(an older version of Urbana 3-body force) does not have this problem. Let us stress again that the repulsive term in the Urbana interaction is isotopically invariant. 
Thus its corrections would affect normal nuclei, symmetric nuclear matter etc.

In conclusion let us give a simple explanation why the collapse takes place (this explanation was proposed by one of our colleagues). Put $n=N/2$ neutrons in one blob, and the other $n=N/2$ 
particles in a second blob, both of volume $V$ very small. A free gas wave function in each blob would give you the kinetic energy $n$ 
times the $E_F$ (Fermi energy), or proportional to $n^{5/3}$. Since $T(r)$ and $V(r)$ in Urbana and Illinois become small for small
$r$, for small blobs, the only three body interaction is when 1 particle
of a triplet is in one blob, and the other two are in the second blob. Pick a distance between the blobs where this three-body interaction is
attractive. You then get $n^3$ triplets, multiplied by this attractive
interaction. The Argonne potential has bounded pair interactions, which contribute as $n^2$ but it also has $L^2$ terms. These are proportional to $p^2$, 
or for small blobs, $p_F^2$ (Fermi momentum), i.e.
proportional to $n^{2/3}$. There are $2n^2$ of these terms, so this gives
an order $n^{8/3}$ bound. The other terms have smaller exponents. Since $n^{8/3} < n^3$ for large $n$, you can always make the triplet term dominate, which leads 
to the collapse of the system.

\end{document}